\documentclass[english]{article}
\usepackage[latin9]{inputenc}
\usepackage{geometry}
\usepackage{float}
\usepackage{amsbsy}
\usepackage{amstext}
\usepackage{amssymb}
\usepackage{graphicx}
\usepackage{setspace} 
\makeatletter
\newcommand{\lyxaddress}[1]{
\par {\raggedright #1
\vspace{1.4em}
\noindent\par}
}

\usepackage{siunitx}
\usepackage{braket}
\usepackage{inputenc} 
\usepackage{xcolor}
\usepackage{upgreek}
\usepackage{textcomp}

\DeclareSIUnit\permille{\text{\textperthousand}}

\makeatother

\usepackage{babel}
\begin{document}

\title{Coherence of a charge stabilised tin-vacancy spin in diamond}

\author{Johannes G\"{o}rlitz$^{1*}$, Dennis Herrmann$^{1}$, Philipp Fuchs$^{1}$,
Takayuki Iwasaki$^{2}$, \\
Takashi Taniguchi$^{3}$, Detlef Rogalla$^{4}$, David Hardeman$^{5}$,
Pierre-Olivier Colard$^{5}$, \\
Matthew Markham$^{5}$, Mutsuko Hatano$^{2}$, Christoph Becher$^{1*}$}
\maketitle

\lyxaddress{$^{1}$Fachrichtung Physik, Universit\"{a}t des Saarlandes, Campus E2.6,
D-66123 Saarbr\"{u}cken, Germany \\
$^{2}$Department of Electrical and Electronic Engineering, Tokyo
Institute of Technology, Meguro, Tokyo 152-8552, Japan\\
$^{3}$International Center for Materials Nanoarchitectonics, National
Institute for Materials Science, 1-1 Namiki, Tsukuba 305-0044, Japan\\
$^{4}$RUBION, Ruhr-Universit\"{a}t Bochum, Universit\"{a}tsstra{\ss}e 150, D-44801
Bochum, Germany\\
$^{5}$Element Six Global Innovation Centre, Fermi Avenue, Harwell
Oxford, Didcot, Oxfordshire, OX11 0QR, United Kingdom }
Email: j.goerlitz@physik.uni-saarland.de; christoph.becher@physik.uni-saarland.de

\begin{abstract}
Quantum information processing (QIP) with solid state spin qubits
strongly depends on the efficient initialisation of the qubit's desired
charge state. While the negatively charged tin-vacancy ($\text{SnV}^{-}$)
centre in diamond has emerged as an excellent platform for realising
QIP protocols due to long spin coherence times at liquid helium temperature
and lifetime limited optical transitions, its usefulness is severely
limited by termination of the fluorescence under resonant excitation.
Here, we unveil the underlying charge cycle, potentially applicable to all group IV-vacancy (G4V)
centres, and exploit it to demonstrate highly efficient and rapid initialisation
of the desired negative charge state of single $\text{SnV}$ centres
while preserving long term stable optical resonances. In addition to investigating the optical coherence, we
all-optically probe the coherence of the ground state spins by means
of coherent population trapping and find a spin dephasing time of
\SI{5(1)}{\mu s}. Furthermore, we demonstrate proof-of-principle
single shot spin state readout without the necessity of a magnetic
field aligned to the symmetry axis of the defect.
\end{abstract}

\section*{Introduction}

Colour centres in diamond have recently emerged as competitive platforms
in the field of QIP \cite{Hanson3NVNetwork,LukinQuantumNetwork,LukinMemoryEnhComm}.
In particular the G4V centres combine long spin coherence times \cite{AtatureQuantumControlSnV,LukinMillikelvin}
with favourable optical properties such as large Debye-Waller factors
\cite{NJPSpecInv,NeuNJP} and transform limited resonance linewidths
\cite{NJPSpecInv,EnglundTransformLimSnV,JelezkoCloseToFourierLimSiV,LukinGeVCloseToFourier}.
While the large spectral diffusion for the nitrogen vacancy centre
in diamond \cite{EnglundOpticalCoherenceNV} is to be expected, G4V
centres should be protected by their inversion symmetry from large
first order Stark shifts. Nevertheless, spectral diffusion due to
second order effects is still common to impose a strong limit on the
usefulness in quantum information and simulation applications \cite{EnglundTransformLimSnV,BeausoleilNVSpecDiff,EnglundStarkEffectSnV,VuckovicElectricalTuningSnV,LukinQuantumRegister,VuckovicSpecDiffSnV},
which makes overcoming these spectral instabilities a crucial task.
The origin of the unstable resonance lines is oftentimes introduced
by fluctuating charges resulting from impurities or lattice defects
in the vicinity of the colour centre. Furthermore, these environmental
impurities and defects can act as charge traps or electron donors
leading to an ill-defined charge state of the emitter. Among the G4V
centres, the $\text{SnV}^{-}$ centre \cite{NJPSpecInv,HatanoSnV,OliveroSnV,GaliGroupIVVac}
shows great promise in terms of spin coherence at easily accessible
liquid helium temperatures \cite{AtatureQuantumControlSnV} and lifetime
limited optical transitions \cite{NJPSpecInv,EnglundTransformLimSnV}.
Unfortunately, its potential for application in QIP is suffering strongly
from the termination of fluorescence under resonant excitation \cite{AtatureQuantumControlSnV,NJPSpecInv,EnglundTransformLimSnV}.
In this work, we investigate the mechanism leading to the charge instability
of a single $\text{SnV}^{-}$ centre in diamond and find it to be
a single photon process. We furthermore fully explore the charge cycle
of the SnV centre and use it to implement a straightforward, highly
efficient and rapid initialisation of the negative charge state while preserving
long-term stablility of dipole allowed optical transitions. We find
that the principle of this charge cycle can also directly be applied
to other G4V centres for efficient control of their charge state.
Subsequently, we probe the coherence of the charge stabilised $\text{SnV}^{-}$
centre by an all-optical coherent population trapping (CPT) scheme,
yielding a spin dephasing time of the ground state of $T_{2}^{*}$=
\SI{5(1)}{\mu s}. By employing a proof-of-principle single shot readout
protocol, we demonstrate the presence of highly cycling optical transitions
even in the case of large angles between the magnetic field and the
symmetry axis of the defect.

\section*{Results}

\subsubsection*{Charge cycle of the $\boldsymbol{\text{SnV}}^{\boldsymbol{-}}$centre}

We investigate single $\text{SnV}^{-}$centres in diamond created
upon ion implantation and subsequent high-pressure-high-temperature (HPHT) annealing (sample NI58, see method section for details on the sample
and the experimental setup) at temperatures of \SI{2}{K}, for which
termination of fluorescence is reliably observed under resonant excitation
of the defect \cite{AtatureQuantumControlSnV,NJPSpecInv,EnglundTransformLimSnV}.
Without additional measures, we find this termination to be permanent,
even after hour long waiting times in the dark. Furthermore, by performing
resonant excitation scans over a large spectral range of several GHz,
we can exclude that the resonance line is simply shifted due to spectral
diffusion, the range of which is typically limited to $\ll$ \SI{1}{GHz}
for the $\text{SnV}^{-}$ centre. Applying an additional \SI{532}{nm}
laser pulse, as often used in experiments with G4V centres, suffices
to recover the fluorescence, but usually leads to a resonance line
shifted by up to several hundred MHz \cite{NJPSpecInv}, much larger
than the typical lifetime-limited linewidths of 20-\SI{30}{MHz}.
A similar effect was also found for germanium vacancy centres in diamond
\cite{GaoGeVOpticalGating}. 
\begin{figure}[H]

\includegraphics[width=1\textwidth]{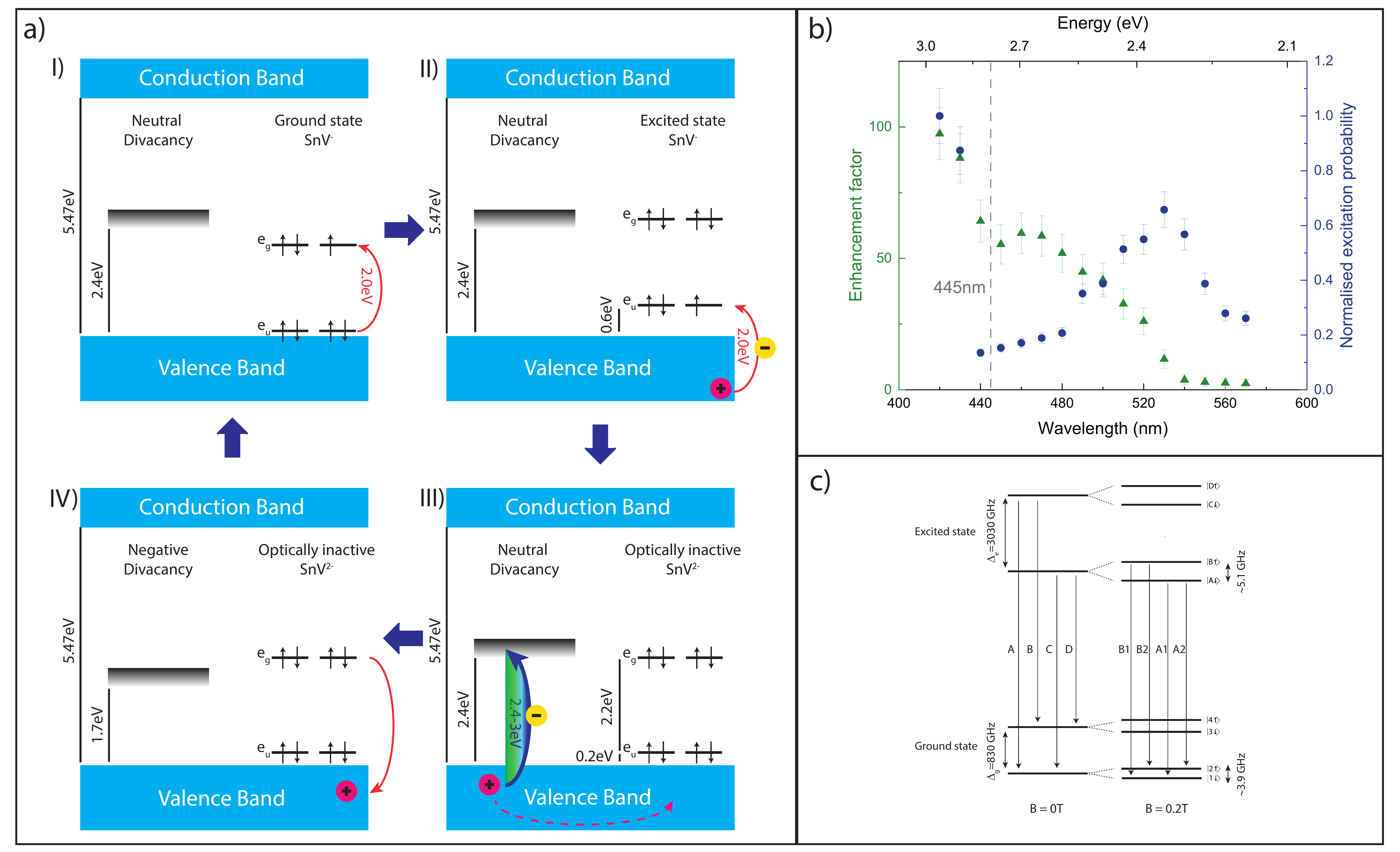}

\caption{\textbf{\label{fig:ChargeProcesses}Charge processes of the SnV$^{-}$centre:
a) }Proposed scheme for the charge cycle of the SnV centre. In I)
the SnV$^{-}$centre gets excited by a resonant \SI{2}{eV} photon.
In II) a second \SI{2}{eV} photon promotes an electron from the valence
band to the SnV$^{-}$centre, transforming it into its optically inactive
charge state SnV$^{2-}$. In III) a photon with an energy above \SI{2.4}{eV}
drives the charge transition of a defect in the diamond lattice, most
likely the neutral divacancy. In this process a hole is created in
the valence band which diffuses in IV) towards the SnV$^{2-}$centre,
where a recombination takes place. All absolute energy positions related
to the charge states of the SnV$^{-}$centre are extracted from \cite{GaliGroupIVVac}.
\textbf{b) }Enhancement of fluorescence factor of an ensemble of SnV$^{-}$centres,
excited resonantly on the C-transition and under addition of a second
laser with tunable wavelength. The enhancement factor is defined as
the ratio between counts with and without the second laser. For comparison,
the normalised excitation probability measured via excitation by the
second laser only is plotted. The excitation probability peaks at
\SI{2.4}{eV} due to excitation to a higher excited state of the SnV$^{-}$
\cite{NJPSpecInv} and an indication of an even higher lying excited
state is seen at photon energies larger \SI{2.8}{eV} most likely
resulting from a deeper lying electron state in the valence band \cite{GaliGroupIVVac}. \textcolor{black}{y-errors (s.d.): Poisson distributed count rate errors.} 
\textbf{c) }Fine structure and Zeeman splitting of the SnV$^{-}$centre.
The relevant optical transitions and splittings at $\text{B}$=\SI{0}{T}
and $\text{B}$=\SI{0.2}{T} for this work are labeled.}
\end{figure}
These spectral shifts impose a strong
limitation on the usefulness of the $\text{SnV}^{-}$centre in QIP.
The origin of the instability of the resonance line is most likely
caused by an altered charge environment after the application of the
\SI{532}{nm} laser. Understanding these charge dynamics is crucial
in order to fully exploit the superior spin coherence of single $\text{SnV}^{-}$centres
at liquid helium temperatures compared to other G4V centres \cite{AtatureQuantumControlSnV,LukinMillikelvin,HatanoSnV,BecherMillikelvin,JelezkoGeVCoherentControl}. 

In order to stabilise the desired negative charge state of the SnV
centre without inducing spectral shifts, we systematically vary the
wavelength of an additionally applied light field. To this end we
perform a measurement in which we use a continuous wave dye laser
to resonantly excite an ensemble of $\text{SnV}^{-}$centres on the
C-transition (see Fig.\ref{fig:ChargeProcesses}c)), add a second
light field provided by a pulsed supercontinuum laser source at a repetition
rate of \SI{80}{MHz} and monitor the count rate on the phonon sideband
(Bandpass filter: \SI{655(47)}{nm}). We scan the wavelength of the
second laser from \SI{420}{nm} up to \SI{580}{nm} with a spectral
bandwidth of \SI{10}{nm} while keeping the power fixed and compare
the count rate of the purely resonant excitation to the count rate
of the two-colour excitation while substracting the counts caused
by the second light field alone (Fig.\ref{fig:ChargeProcesses}b)).
For wavelengths of the second light field $\gtrsim$\SI{520}{nm}
the ensemble fluorescence remains at a low level due to the fluorescence
termination discussed above. However, we observe a fluorescence enhancement
starting at wavelengths (energies) shorter (larger) than \SI{520}{nm}
(\SI{2.4}{eV}), for which we define the enhancement factor $\beta=\frac{I_\text{w}}{I_\text{wo}}$ as the
ratio between emitted fluorescence intensity with ($I_\text{w}$) and without ($I_\text{wo}$) the second laser being applied. 
This is depicted by the green triangles in Fig.\ref{fig:ChargeProcesses}b). The observed
fluorescence enhancement indicates a stabilisation of the desired
negative charge state for photon energies above \SI{2.4}{eV}, hinting
at a charge transfer process occuring at a certain threshold. This
result is confirmed by a similar measurement on a second HPHT annealed sample labelled BOJO\_001,
see method section and Supplementary Note 3 for further information. \newline Before devising a model for the SnV charge
cycle, we further investigate the charge dynamics. To this end, we
select \SI{445}{nm} as the optimised wavelength for recovery of the
negative charge state of the SnV centre, because it lies at the optimum
of fluorescence enhancement while minimising the direct excitation
of the $\text{SnV}^{-}$centre. The latter can be seen by the normalised excitation probability, 
which is depicted as blue dots in Fig.\ref{fig:ChargeProcesses}b), as function of the excitation wavelength. 
We derive it from the count rate emitted upon excitation with the second wavelength laser only and normalise
 it by the maximum count rate measured for all excitation wavelengths. The minimised direct excitation is important, since the termination of 
the fluorescence is only occuring from the excited state of the $\text{SnV}^{-}$centre as we will show below. Thus direct excitation 
with the charge initialisation laser can lead to a subsequent termination of the fluorescence stimulated by a 
second photon from the initialisation laser and therefore to a reduced initialisation efficiency. 
We now evaluate the number of photons
that are taking part in the charge transfer process under resonant
excitation. To this end, we employ a \SI{10}{ms} laser pulse at \SI{445}{nm}
to initialise the negative charge state of a single $\text{SnV}^{-}$centre
(Emitter 1, characterisation of the single photon emission can be
found in the Supplementary Note 1). Subsequently, we apply a \SI{1}{s}
long laser pulse resonant with the C-transition of the defect and
collect photons emitted into the phonon sideband. The duration of
the resonant laser pulse is chosen sufficiently long such that charge
transfer will occur during this time even for the lowest power used.
\begin{figure}[H]

\includegraphics[width=1\textwidth]{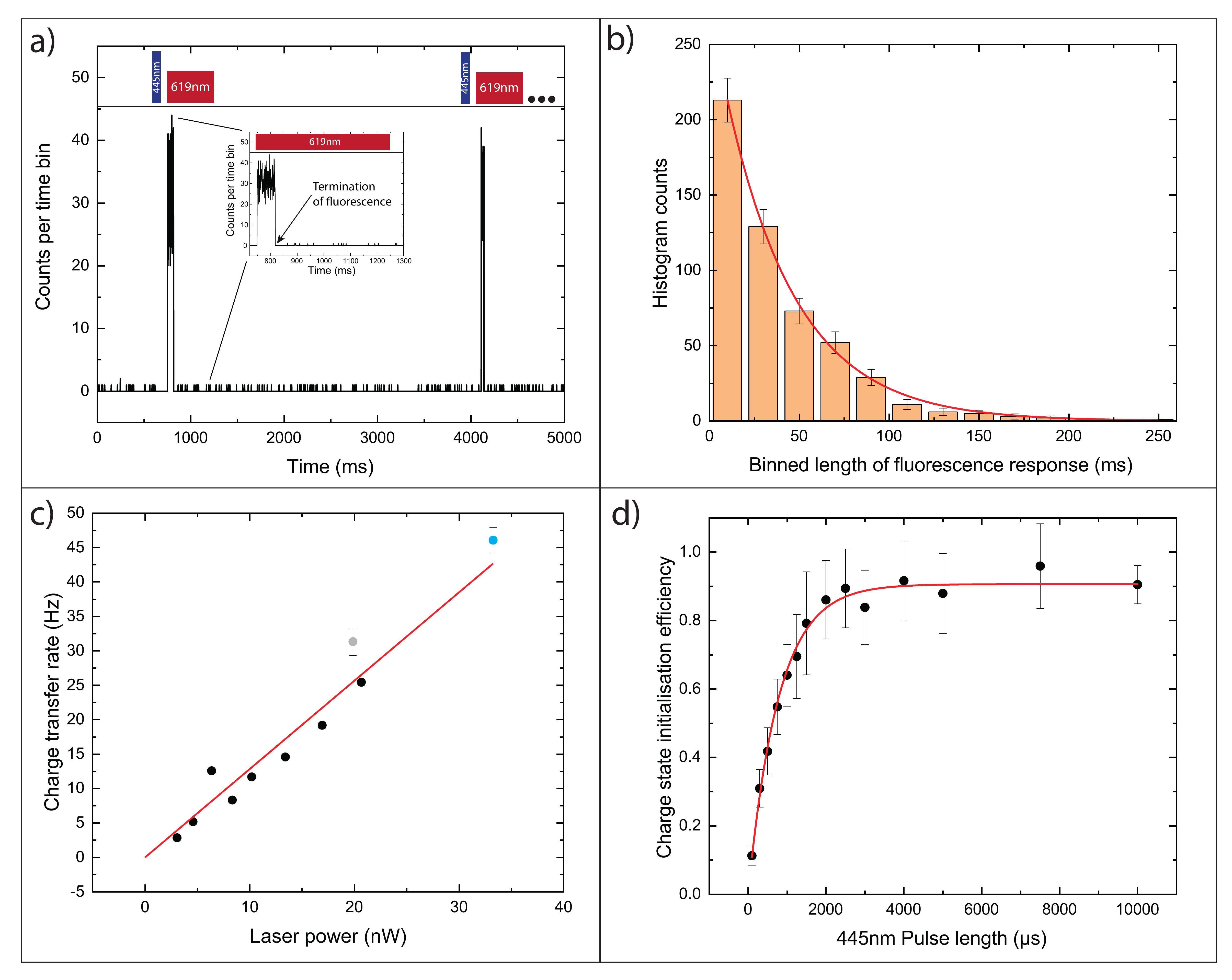}

\caption{\textbf{\label{fig:ElectronCaptureAndInitialisation}Electron capture
and charge initialisation of a single SnV$^{-}$centre: a) }Exemplary
part of a measurement sequence of the charge initialisation and electron
capture experiments on a single SnV$^{-}$centre (Emitter 1). After
the \SI{10}{ms} laser pulse at \SI{445}{nm} initalises the charge
state, a fluorescence response by the \SI{1}{s} long resonant laser
pulse on the C-transition is visible. The fluorescence response lasts
until electron capture occurs. \textbf{b) }Histogram of the lengths
of the fluorescence responses for a given resonant laser power of
\SI{20.7}{nW}. An exponential fit yields the charge state lifetime
until charge transfer occurs. This lifetime translates directly to
an charge transfer rate for every given laser power. \textcolor{black}{y-errors (s.d.): Poisson distributed count rate errors.} \textbf{c) }Charge
transfer rate plotted against the resonant excitation laser power.
The linear dependence of the charge capture rate on the power indicates
a single photon process from the excited state, since termination
of fluorescence does not occur, if the laser is not resonant with
the C-transition. The blue data point corresponds to the laser power
chosen for the charge state initialisation measurement in d), while
the greyed out data point results from a measurement aborted earlier
than the others. \textcolor{black}{y-errors (s.d.): Fitting error from measurements in \textbf{b)}.} \textbf{d) }Measurement of the charge state initialisation
efficiency. \textcolor{black}{The \SI{445}{nm} laser pulse of variable pulse length
is followed by a \SI{0.5}{s} long resonant laser pulse with a power
chosen such, that electron capture occurs with a probability close to unity
for each pulse.} The initialisation efficiency is extracted by counting
the number of fluorescence response pulses that occur and dividing
them by the total number of pulses. \textcolor{black}{For increasing pulse length of
the blue laser, the initialisation efficiency follows an exponential
law with a time constant of $\tau_{\text{CI}}=$\SI{780(27)}{\micro\second} and saturates at \SI{91(1)}{\%}.} \textcolor{black}{ y-errors (s.d.): Poisson distributed count rate errors.}}
\end{figure}
An exemplary excerpt of such a time trace is depicted in Fig.\ref{fig:ElectronCaptureAndInitialisation}a).
We repeat this measurement about 500 times for each laser power and
extract the charge state lifetime, which we define as the duration
until charge transfer occurs under resonant excitation, by evaluating
the duration of emission of the $\text{SnV}^{-}$centre until the
fluorescence is terminated. A histogram of such a measurement together with a mono-exponential fit is shown in
 Fig.\ref{fig:ElectronCaptureAndInitialisation}b. The charge lifetime is extracted as the time constant of the exponential decay. 
The inverse of the lifetime yields the charge transfer rate, which
increases linearly with the resonant excitation power (see Fig.\ref{fig:ElectronCaptureAndInitialisation}c)). 
The linear fit in red, with the y-intercept being set to zero, yields a slope of $1.28(5)\frac{\text{Hz}}{\text{nW}}.$
We therefore conclude that charge transfer occurs as a single photon
process. It takes place when the $\text{SnV}^{-}$centre is in the
excited state, since termination of fluorescence does not occur when
the excitation laser is far detuned from resonance. In such a case
one would intuitively expect a two photon process where the first
photon excites the emitter and the second photon causes the charge
transfer, therefore leading to a quadratic power dependence of the
charge transfer rate. However, the charge transfer happens at time
scales much longer (6 orders of magnitude) than the lifetime of the
excited state. We thus expect that the charge transfer starts from
a steady state population of the excited state created by the resonant
excitation and consequently involves only a single photon.

To explain the charge transfer processes ocurring under resonant excitation
and illumination with a second light field we start with the position
of the energy levels of the $\text{SnV}^{-}$ centre in the diamond
bandgap (Fig.\ref{fig:ChargeProcesses}a)). All absolute energy values of SnV states are taken
from theoretical calculations conducted in \cite{GaliGroupIVVac}. Under resonant excitation with the ZPL energy of about \SI{2.0}{eV}, 
the absolute energy position of the $\text{SnV}^{-}$ levels shift about \SI{0.6}{eV} upwards in the diamond bandgap, which is \SI{5.47}{eV} wide. The highest
occupied levels of the $\text{SnV}^{-}$ excited state thus lie \SI{0.6}{eV}$+$\SI{2.0}{eV}
above the valence band and consequently \SI{5.47}{eV}$-$\SI{2.6}{eV}$=$\SI{2.87}{eV}
below the conduction band. \\
We exclude a charge transfer process under resonant illumination by
photo-ionisation of an electron from the excited state to the conduction
band based on the following facts and observations: (i) the photon
energy of \SI{2.0}{eV} is not sufficient for a one photon-excitation
from the excited state to the conduction band; (ii) a two-photon process
from the excited state ($2\cdot$\SI{2.0}{eV}$>$\SI{2.87}{eV}) is
excluded as we observe a one-photon charge transfer process (see discussion
above); (iii) electron ionisation would leave the SnV center in its
neutral state. Although it is very challenging to unambiguously observe
emission from different charge states (due to potential charge transfer
happening upon excitation) we gather evidence against transfer into
the neutral charge state: When exciting an ensemble of $\text{SnV}^{-}$
centres resonantly without any charge stabilising laser being applied
and measuring spectra integrated longer than \SI{10}{min} in the
spectral region from 620-\SI{900}{nm}, these measurements do not
show the slightest hint of $\text{SnV}^{0}$ centre emission, which
is theoretically predicted to lie at \SI{680(40)}{nm} \cite{Gali_NeutralG4V}. 
In this case, the \SI{620}{nm} excitation should not only transfer
the SnV centres to their neutral charge state but also have significant overlap with its
excited state vibrational ladder. For comparison, the vibrational ladder spans about \SI{100}{nm} for the $\text{SnV}^{-}$ \cite{NJPSpecInv}. Thus, the \SI{620}{nm} radiation should also excite the $\text{SnV}^{0}$, which we do not observe.
Transfer of a single charge then only leaves the alternative of transfer
to the doubly negative charge state $\text{SnV}^{2-}$ which is an
optically inactive state and cannot be observed spectroscopically.
However, our combined observations as discussed above lead to the
conclusion that the dark state of the SnV centre is indeed the $\text{SnV}^{2-}$.
This claim furthermore is in agreement with recent experimental findings
on the $\text{SiV}^{-}$ center where a doubly negative charge state
was identified as the dark state \cite{KolkowitzLocalChargeEnvironment}. 

We here go a step further and propose a model which not only explains
the SnV centre charge dynamics but should be generally valid for all
G4V centers. Our model is based on the charge state conversion process
identified in \cite{KolkowitzLocalChargeEnvironment}, i.e. capture
of an electron from the valence band (hole generation) for the $\text{SnV}^{-}$
$\rightarrow$ $\text{SnV}^{2-}$ process and release of an electron
to a hole in the valence band (hole capture) for going from $\text{SnV}^{2-}$
$\rightarrow$ $\text{SnV}^{-}$. The latter process requires the
presence of a valence band hole in the vicinity of the SnV center
which might be created by another close-by defect upon photo-excitation.
We identify this defect by the measurement of fluorescence enhancement
presented in Fig.\ref{fig:ChargeProcesses}b): The onset and also
the shape of the fluorescence enhancement curve agree very well with
the theoretically proposed \cite{Gali_DiVacancyChargeTransition}
and experimentally confirmed \cite{Dannefaer_ChargeTransitionDiVacancy}
charge transition of the neutral divacancy in diamond to its negative
charge state. This charge transition occurs due to promotion of an
electron from the valence band to the divacancy \cite{Gali_DiVacancyChargeTransition,Dannefaer_ChargeTransitionDiVacancy}.
It was also reported, that the efficiency of this ionisation reaches
unity charge conversion at photon energies \textgreater{}\SI{3}{eV}.
The diamond sample investigated in this study is very likely to contain
many divacancies even after the HPHT annealing (\cite{HPHTBook}, p.32-34), since
the implantation of one tin ion with an energy of \SI{700}{keV} creates about 3000 vacancies (Monte Carlo Simulation, SRIM)
which tend to form divacancies in an exothermic reaction during annealing
due to single vacancies becoming mobile \cite{DivacancyFormation}. Hence, monovacancies will no longer be present as was also found in \cite{Dannefaer_ChargeTransitionDiVacancy}. 
Regarding other common impurities in diamond such as boron and nitrogen, they are present in low concentrations as specified in the methods section. However, their typical absorption energies \cite{Zaitsev} do not match the observed fluorescence enhancement threshold. While we cannot exclude that
the charge transition originates from a different vacancy or impurity complex in diamond, we find the divacancy to be the most likely candidate.
Based on these considerations we propose the following charge cycle
for the $\text{SnV}^{-}$centre: A resonant \SI{620}{nm} (\SI{2}{eV})
photon first excites the $\text{SnV}^{-}$centre to a fully occupied
$\text{e}_{\text{g}}$ state (see Fig.\ref{fig:ChargeProcesses}a)).
A subsequent \SI{620}{nm} photon promotes an electron from the valence
band to the $\text{SnV}^{-}$ centre's empty $\text{e}_{\text{u}}$
orbital and transforms it into its optically inactive charge state
$\text{SnV}^{2-}$. The created hole left in the valence band diffuses to a defect in the vicinity where it is caught, the $\text{SnV}^{2-}$ cannot recombine with it and the $\text{SnV}^{-}$ fluorescence is terminated. A photon with a wavelength (energy) of 420-\SI{520}{nm}
(2.4-\SI{3}{eV}) excites an electron from the valence band to a defect
in the diamond lattice, the most likely candidate being the neutral
divacancy. This process creates a hole in the valence band which diffuses
towards the $\text{SnV}^{2-}$ centre where a recombination leads
to the return to $\text{SnV}^{-}$, which closes the cycle. In the second run of the charge cycle the divacancy is in its negative charge state and will act as the trap for the hole created in step II in Fig.\ref{fig:ChargeProcesses}a). For our considerations above, 
we restricted the charge repump energy to the range of 2.4-\SI{3}{eV}, since we were limited to energies
 below \SI{3}{eV} by the utilised supercontinuum laser in the measurement of the enhancement factor. 
It might well be possible that higher photon energies are also suitable for charge stabilisation of the $\text{SnV}^{-}$. The developed charge cycle scheme
is also able to explain the large spectral shifts that occur when
applying \SI{532}{nm} laser light as a charge repump. The energy
of the green light is not sufficient to efficiently convert all the surrounding
divacancies to their negative charge state and therefore leaves a
fluctuating charge environment. The \SI{445}{nm} light, in contrast, will also be sufficient to ionise most common impurities that might occur in low densities (B,N,..) and therefore create a charge environment that is the same after each application of the charge stabilisation laser. A direct comparison of the effect of blue and green laser light charge stabilisation can be found in the Supplementary Note 6.

Building on these results, we evaluate the efficiency with which we
can initialise the negative charge state of the investigated $\text{SnV}$
centre. For this we apply a \SI{445}{nm} initialisation pulse with
variable pulse length and a cw power of \SI{50}{\mu W} which is followed
by a \SI{0.5}{s} resonant laser pulse creating a fluorescence response
if the $\text{SnV}^{-}$ charge state is properly initialised. A timegap of a few microseconds ensures that the effects of charge stabilisation and electron capture do not overlap. \textcolor{black}{The
power of the resonant laser pulse is chosen corresponding to an electron
capture rate of \SI{46(2)}{Hz} (blue dot in Fig.\ref{fig:ElectronCaptureAndInitialisation}c))
ensuring electron capture with a probability close to unity. The theoretical value can be calculated to $1-\exp(-0.5\text{s}\cdot46\frac{1}{\text{s}})$, with the error to unity being below $10^{-10}$ under consideration of the error bar of the electron capture rate.}
For each blue pulse length, we repeat the measurement about 250 times
and count the number of initialisation pulses after which a fluorescence
response is observable. The ratio of observed pulses to the total
number of pulses used yields the initialisation efficiency, which
is depicted in Fig.\ref{fig:ElectronCaptureAndInitialisation}d).
\textcolor{black}{We reach a saturation value of \SI{91(1)}{\%} charge initialisation,
which we fit by a mono-exponential growth law, $1-\exp(-\frac{t}{\tau_{\text{CI}}})$, with $\tau_{\text{CI}}$ being
 the time constant of the charge repumping. The time dependence of the pumping results from transferring the
electron from the valence band to the neutral divacancy and the subsequent
hole capture of the $\text{SnV}^{2-}$. Its dependence on the blue laser power is discussed in the Supplementary Note 6. We find that the rate of charge state initialisation increases linearly (190(8)$\frac{\text{Hz}}{\upmu\text{W}}$) with blue laser power, again confirming the model of a single-photon charge transfer process. At the largest laser power used (\raisebox{-0.9ex}{\~{}}\SI{600}{\micro\watt}), we demonstrate rapid charge initialisation in about \SI{10}{\micro\second}. \SI{4(1)}{\%} of the missing initialisation
efficiency are due to the number of pulses where electron capture
occurs too fast to create a sufficiently strong fluorescence response
to be distinguished from the dark counts and background in our system.
The missing \SI{5}{\%} to unity might be caused either by the restrictive threshold on fluorescence counts
 that we set for a positive pulse count after initialisation or by spurious electron
capture by the blue laser itself due to the small but measurable direct excitation probability and the possibility of a second blue laser photon transferring an electron from deep within the valence band to the  $\text{SnV}^{-}$, thus turning it dark.} 
It is worth noting, that we found
this concept of initialisation to work on every emitter that we investigated
in this sample and a similar HPHT annealed second sample (see Supplementary Note 3). Detailed results of a second emitter under investigation
reaching \SI{97}{\%} of charge state initialisation efficiency can
be found in the Supplementary Note 6.\newline \textcolor{black}{Another noteworthy finding is that under resonant pulsed excitation well below saturation with a low duty cycle (about \SI{0.2}{\permille}) in order to prevent electron capture, the charge state set by a one-time application of the initialisation laser was found to persist over more than an hour of measurement time. In the dark it was maintained over several hours and thus we conclude that its lifetime is only limited by laser induced electron capture. While we utilise pulsed and continuous wave stabilisation schemes within this work, generally it would be fully sufficient to employ pulsed charge stabilisation with high efficiency and repeat initialisation with a rate larger than the electron capture rate imposed by the resonant excitation.} Further results on the
photophysics of $\text{SnV}^{-}$ centres in low-pressure-low-temperature
annealed diamond samples which are induced by the charge cycle described
above will be published elsewhere.

We would furthermore like to emphasise that the proposed charge cycle
model and and the method for stabilising the negative charge state
are most probably universal to the G4V centers in diamond. The reason
is that their general electronic level structure is identical and
the lower $\text{e}_{\text{u}}$ orbitals generally lie close to the
valence band edge such that the charge transfer as discussed here
for SnV centres works in an analogous fashion. Small differences might
evolve due to different concentrations of divacancies introduced in
the implantation process and absolute energy positions within the
band gap of diamond. This claim is strongly supported by an experiment
on a resonantly excited ensemble of $\text{SiV}^{-}$centres in which
a very similar fluorescence enhancement under illumination with a
second tunable laser is observed (see Supplementary Note 4) and furthermore
by the findings reported in \cite{KolkowitzLocalChargeEnvironment}.

\subsubsection*{Long term stability of the resonances of a charge stabilised SnV$^{-}$}

\begin{figure}[H]

\includegraphics[width=1\textwidth]{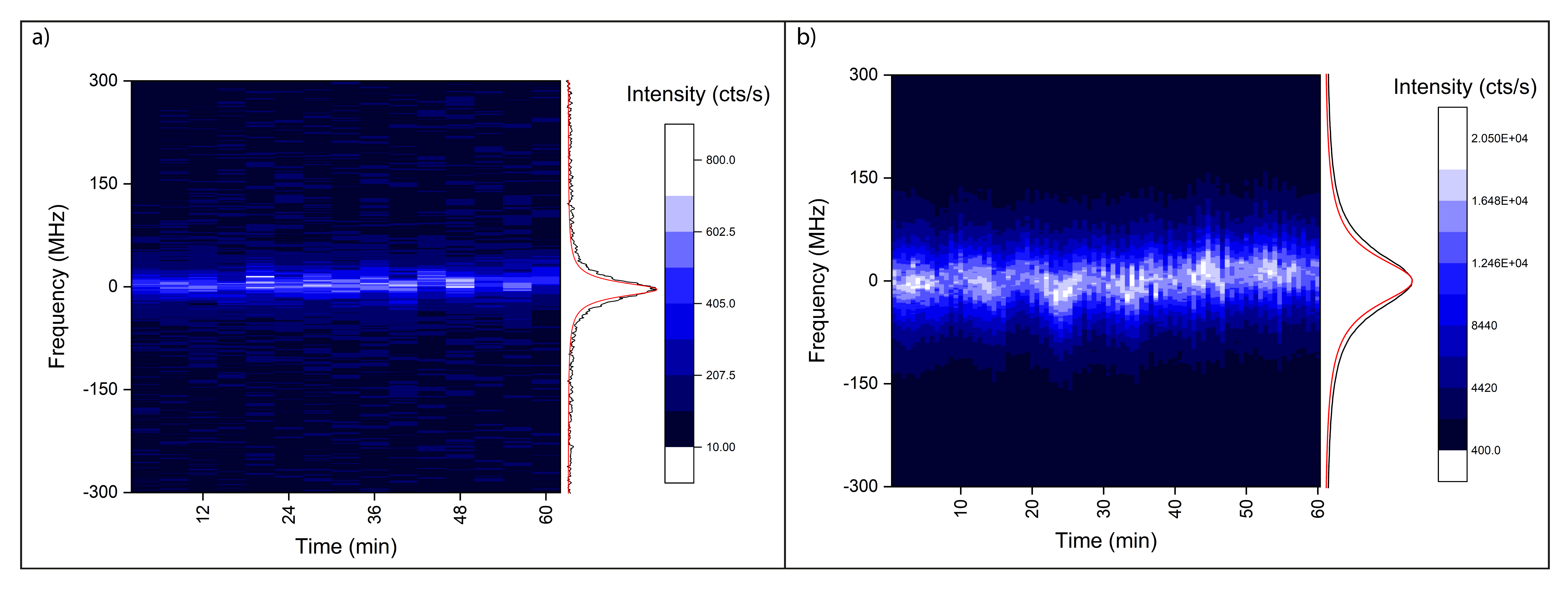}

\caption{\textbf{\label{fig:LongtermStability}\textcolor{black}{Long-term} stability of the C-transition
of SnV$^{-}$centres: a) }Low excitation power of \SI{0.5}{nW} \textcolor{black}{long-term}
PLE in order to resolve spectral shifts caused by the charge environment
or the \SI{445}{nm} laser without initiating the charge cycle of
the centre. The constantly charge stabilised emitter's line position
exhibits a standard deviation of less than \SI{4(2)}{MHz} over one
hour of measurement. On the right, the integrated spectrum over all
scans (black) with a width of \SI{33(2)}{MHz} is compared with the
Fourier limited Lorentzian line (red, \SI{25}{MHz}) \textbf{b) }\textcolor{black}{Long-term}
PLE on Emitter 2 with a power of \SI{10}{nW} ensuring electron capture
in every resonance scan while minimising power broadening. In the
duration of one hour, the central resonance position exhibits a standard
deviation of \SI{10(2)}{MHz}. The oscillations with a period of about
\SI{10}{min} in the central resonance position are related to the
air conditioning cycle in our laboratory. On the right, the integrated
spectrum over all scans (black) with a width of \SI{103(2)}{MHz}
is compared to a Lorentzian that is only subject to power broadening
(red) and exhibits a width of \SI{88(2)}{MHz}.}
\end{figure}

The ability to reliably initialise the negative charge state of the
$\text{SnV}^{-}$centre enables us to investigate the influence of
spectral diffusion on the long term stability of the electronic transitions.
\textcolor{black}{There are two cases to distinguish: Firstly, light-induced fluctuations
of the charge environment that introduce Stark shifts \cite{EnglundStarkEffectSnV,VuckovicElectricalTuningSnV}
as a result of the charge stabilisation laser being applied. These disturbances occur without the $\text{SnV}^{-}$centre undergoing a full charge cycle (e.g. the negative charge
state is preserved), resulting in a dynamic equilibrium of the charge environment. In the second case, this equilibrium is disturbed when the $\text{SnV}^{-}$centre captures an electron under resonant excitation and the charge cycle is initalised. It is not guarantueed that the hole necessary for the recombination will result from the same divacancy for each charge cycle.  It is thus 
possible that the near scale charge environment will not end up in its
original state.} Therefore, we expect stronger influence on the spectral
diffusion in the latter case. In order to separate the two effects,
we perform photoluminescence excitation (PLE) spectroscopy and use
very low excitation powers below \SI{0.5}{nW} in order to avoid power
broadening and electron capture. \textcolor{black}{The power is chosen such that it is about a factor of four below the saturation power (\raisebox{-0.9ex}{\~{}}\SI{2}{nW}).} In this way, one can resolve very
small spectral shifts on the order of the natural linewidth of about
\SI{25}{MHz} that are only caused by environmental fluctuations.
During the measurement, the $\text{SnV}^{-}$centre is permanently
charge stabilised by continuous wave \SI{445}{nm} radiation. The resonance
line peak position shift exhibits a standard deviation of less than
\SI{4(2)}{MHz} and a total width of the summed PLE spectra of \SI{33(2)}{MHz}
for the whole duration of the measurement, as displayed in Fig.\ref{fig:LongtermStability}a).
This value has an uncertainty of \SI{2}{MHz} caused by the uncertainty
of our wavemeter which was measured independently. Due to a technical
problem of our positioning system within the cryostat, after this
measurement, emitter 1 could no longer be accessed. The following
measurements were therefore performed on a different emitter (Emitter
2, see Supplementary Note 1) on which all
concepts, that were shown before, work in the same manner. On this
$\text{SnV}^{-}$centre, we investigate the effect of the emitter
undergoing the charge cycle by choosing a laser power of \SI{10}{nW}
as a trade-off to ensure continuous electron capture in every single
resonance scan while introducing as little as possible power broadening.
Even in this case we observe only small shifts in the central resonance
position with a standard deviation of \SI{10(2)}{MHz} and a total
width of the summed PLE spectra of \SI{103(2)}{MHz} compared to \SI{88}{MHz}
purely power broadened linewidth, see Fig.\ref{fig:LongtermStability}b).
The small but visible oscillations with a period of about \SI{10}{min}
in the central resonance position are related to the air conditioning
cycle in our laboratory. We consequently claim that the shift is caused
by tiny alterations of the position of the blue laser on the sample
with respect to the emitter, since slightly misaligning the blue laser
spot on the sample has similar effects. This is in agreement with
a spatially asymetric charge distribution caused by the ionised divacancies.
This effect has the potential of a tuning mechanism of \textless{}\SI{100}{MHz}
in resonance frequency, limited by imperfect charge initialisation
for strongly misaligned blue laser light. \textcolor{black}{It is remarkable that contrary
to the observations reported in \cite{VuckovicSpecDiffSnV}, the charge
stabilisation enables us to limit spectral fluctuations to less than
\SI{16}{\%} (\SI{11}{\%}) of the homogeneous (power broadened) linewidth
of \SI{25}{MHz} (\SI{88}{MHz}) over the course of one hour. For the shifts observed in the low power
measurement, even for photons separated by an arbitrary time distance within
the one hour measurement, a Hong-Ou-Mandel visibility close to \SI{90}{\%}
would be achievable for photons emitted from this $\text{SnV}^{-}$centre (calculated using the simulation tool implemented in \cite{BecherLimIndistSSPs} ("remote HOM testbed"), available from the website of the journal). The same holds true when interfering photons retrieved from two emitters with the same properties, which is an important prerequisite for effectively entangling two $\text{SnV}^{-}$centres. As the experiments on the additional emitters presented in Supplementary Notes 3 and 6 demonstrate, finding two emitters similarly marginally impacted by spectral diffusion is straightforward in the investigated samples.} Thus, the photons
emitted by a charge stabilised $\text{SnV}^{-}$centre should be highly
indistinguishable, which is a key factor in many QIP protocols and
quantum communication. \textcolor{black}{Since the measurements for low and high power excitation had to be conducted on two different emitters, 
an additional measurement on a further emitter accessing both power limits is presented in the Supplementary Note 6, emphasising frequency stability of 
$\text{SnV}^{-}$centres obtained by the charge stabilisation scheme presented in this work.}

\subsubsection*{Coherent population trapping}

With the means of a stable $\text{SnV}^{-}$centre resonance line
at hand, we explore the possibility of all-optical coherent manipulation
of its spin degree of freedom. To this end, we perform CPT between
the states $\ket{1\downarrow}$, $\ket{2\uparrow}$ via the excited
state $\ket{\text{A\ensuremath{\downarrow}}}$. In order to lift the
degeneracy of the spin states, we apply a magnetic field $B$=\SI{0.2}{T}
at an angle of \SI{54.7}{\degree} relative to the axis of the defect.
When applying a magnetic field to the $\text{SnV}^{-}$centre, the
splitting of the spin states and the long ground state spin $T_{1}$
time , which we measure to be larger than \SI{20}{ms} at \SI{200}{mT}
and \SI{1.7}{K} (see Supplementary Note 5) lead to a vanishing PLE
signal due to optical pumping. Introduction of a weak population repump
laser at \SI{532}{nm} leads to the recovery of the PLE signal of
the spin conserving (SC) transitions, while the spin flipping (SF)
transitions are so weak, that they can be found only via CPT measurements.
This is following directly from the isolated atom like selection rules
that are a result of the low strain environment and the strong spin-orbit
coupling in comparison with the Jahn-Teller effect present for the
$\text{SnV}^{-}$centre \cite{EnglundTransformLimSnV}. To generate
the light fields for CPT we set the carrier laser frequency resonant
to the SF transition A2 and we generate the second laser field with
the first harmonic of an electro optical phase modulator (EOM). An
exemplary characteristic single scan CPT signal is shown in Fig.\ref{fig:CPT}a),
which is fitted using a three level density matrix formalism (see
Supplementary Note 7). We relate count rate to populations of the
excited state in the following way: In PLE scans with an excitation
power $P\gg P_{\text{sat}}$ and without a magnetic field being applied,
we measure a fluorescence rate of \textasciitilde{}\SI{45}{kcts/s}
into the phonon sideband, which corresponds to the limit of incoherent
excitation and therefore relates to \SI{50}{\%} of the population residing
in the excited state. This gives us an estimate for the relation of
measured count rate to population. A more exact value under the given
experimental setting is derived by conducting a high power CPT measurement,
shown in Fig.\ref{fig:CPT}a) and fitting it with our model. The mismatch
of the simulation and the data on the high frequency side of the measurement
is caused by approaching the edge of the amplification range of the
microwave amplifier used, therefore effectively changing the power
in the sideband during the scan. We subsequently decrease the laser
power stepwise and change the ratio between control and signal field
via the applied microwave power to the EOM. These sets of measurements
are used to step by step eliminate free parameters in the fit such
as the ratios between Rabi frequencies and laser power, the branching
ratio between SC and SF transitions and the exact number of dark counts
resulting from background fluorescence of the sample induced by the
blue laser light and the dark counts of the avalanche photo diodes
(APD) which can be found in the Supplementary Note 7. Finally the
most interesting parameter, the spin ground state decoherence rate
is fitted and yields a value of \SI{64(10)}{kHz}, see Fig.\ref{fig:CPT}b).
This relates to a spin \textcolor{black}{dephasing} time of $T_{2}^{*}$= \SI{5(1)}{\mu s}
which is slightly larger than found in \cite{AtatureQuantumControlSnV,EnglundTransformLimSnV}.
In comparison to \cite{AtatureQuantumControlSnV}, the branching ratio
$\eta$ between SC and SF transitions amounts to \textasciitilde{}650(100),
which is about a factor 8 larger. This is due to working with an emitter
in a nearly unstrained diamond environment.

\begin{figure}[H]

\includegraphics[width=1\textwidth]{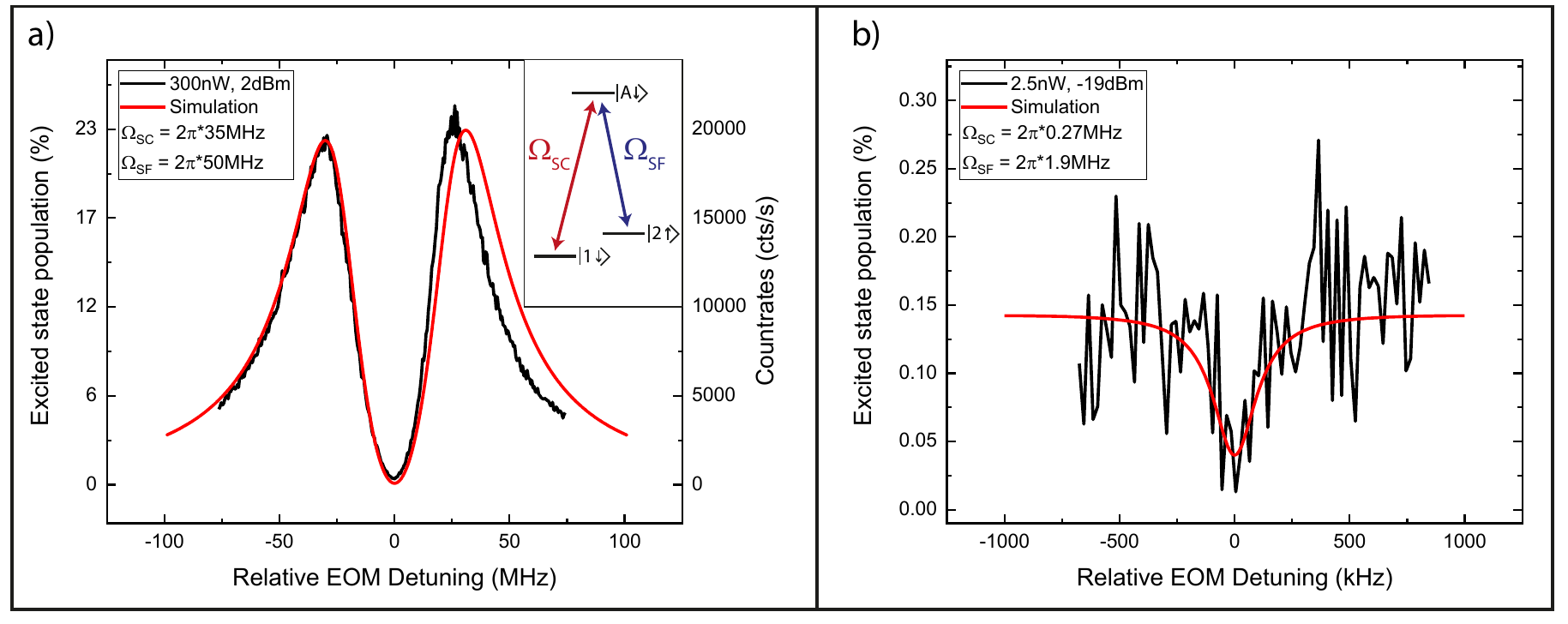}

\caption{\textbf{\label{fig:CPT}CPT measurement on a charge stabilised SnV$^{-}$centre:
a) }CPT measurement on Emitter 2 with \SI{300}{nW} of total laser
power. This measurement is used to relate measured count rate to excited
spin state populations of the $\text{SnV}^{-}$centre. The mismatch
of the simulation and the data on the high frequency side of the measurement
is caused by approaching the edge of the amplification range of the
microwave amplifier used, thereby effectively decreasing the sideband
laser power. \textbf{b)} Measurement of the CPT dip width with lowest
possible total laser power, while preserving a reasonable signal to
noise ratio. The data is fitted with a ground state decoherence rate
of \SI{64(10)}{kHz} relating to a spin dephasing time of $\text{T}_{2}^{*}$=
\SI{5(1)}{\mu s}.}
\end{figure}

\textcolor{black}{We wish to emphasize that coherent optical spin control via CPT on
an unstrained emitter in a clean diamond environment is only possible
using the charge stabilisation method, enabling remarkable long-term stability of the resonance frequency and thus the CPT resonance. In a material where strain is sufficiently large to lower the branching ratio by almost an order or magnitude compared to our findings, CPT with less effective charge stabilisation is possible on selected emitters \cite{AtatureQuantumControlSnV}. We would like to point out that the principles presented in this paper worked straightforwardly on several emitters within our samples, where no preselection of the emitters was applied.}

\subsubsection*{Single shot readout}

\begin{figure}[H]

\includegraphics[width=1\textwidth]{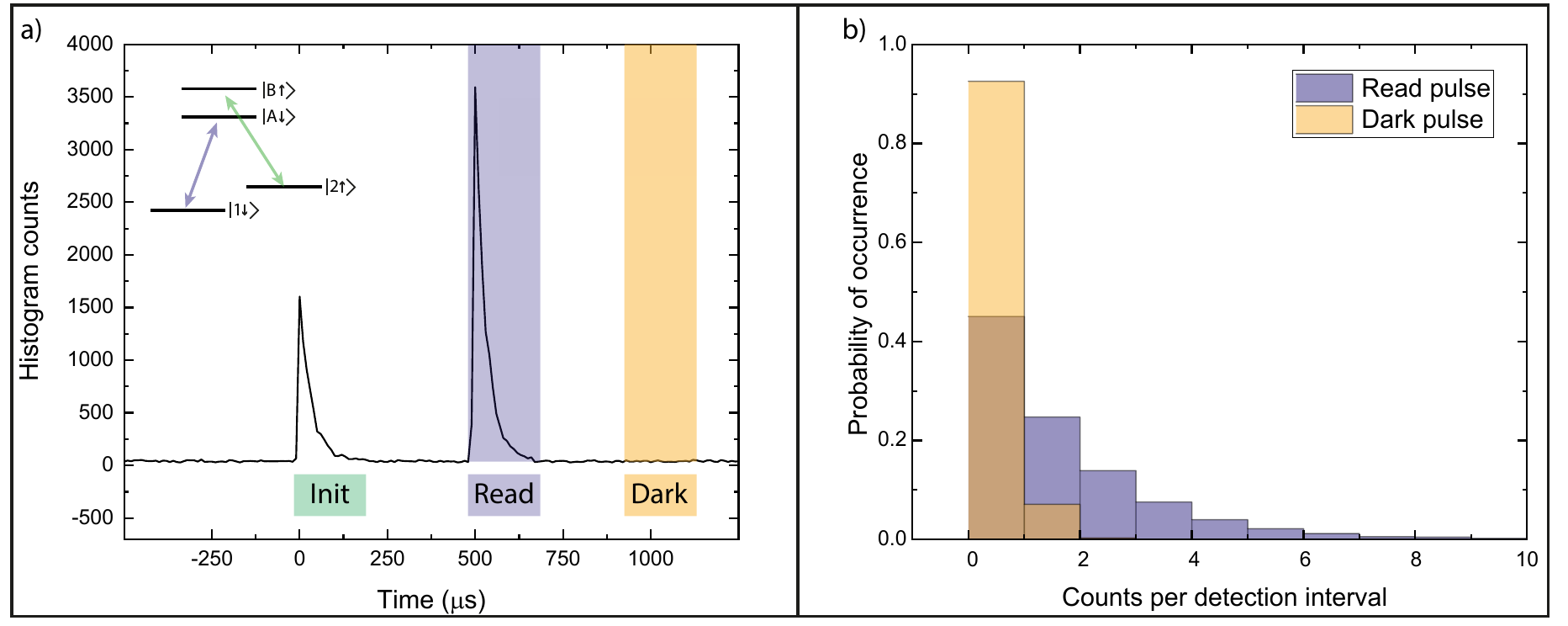}

\caption{\textbf{\label{fig:SingleShot}Single shot readout measurement on
a charge stabilised SnV$^{-}$centre: a) }Histogram of the single
shot readout measurement on Emitter 2. The \SI{200}{\mu s} long initialisation
pulse resonant with the A1 transition initialises the spin with an
efficiency of \SI{98.9(4)}{\%} in the state $\ket{2\uparrow}$. For
each \SI{200}{\mu s} long read pulse we detect an average of 1.21
(1.13) photons without (with) dark count substraction. \textbf{b)
}Histogram of photon counts in the readout time inverval compared
to a time interval of the same length where no laser pulse was applied.
The overlap of the two measurements, while taking the imperfect state
initialisation into account, yields a total readout fidelity of \SI{74}{\%}
when setting 1 photon count as threshold for a state to be detected
as bright. }
\end{figure}

As an outlook to further applications, we investigate whether the
strong selection rules, that are described in the previous section,
lead to sufficiently strong cycling transitions even when the magnetic
field is applied at a large angle with respect to the symmetry axis
of the $\text{SnV}^{-}$centre. To this end, we probe the cyclicity
by investigating the possibility of a single shot readout. \textcolor{black}{The experiment
consists of a \SI{200}{\micro\second} long initialisation pulse resonant
with the SC transition B2 and a \SI{200}{\micro\second} read pulse on the
SC transition A1, separated by \SI{300}{\micro\second} waiting time. In the subsequent waiting time of about \SI{50}{ms} following the two pulses, thermal equilibrium spin population of about \SI{50}{\%}  in each state is approached (spin lifetime of \SI{22}{ms}, see Supplementary Note 5). Therefore, the initialisation peak in  Fig.\ref{fig:SingleShot}a) amounts to roughly half the height of the read out peak.} As it can be seen in Fig.\ref{fig:SingleShot}a),
we achieve an initialisation efficiency of \SI{98.9(4)}{\%} in state
$\ket{2\uparrow}$ while detecting an average of 1.21 (1.13) photons
per pulse without (with) dark count substraction. This number is derived
from comparing the photons collected in the \SI{200}{\mu s} read
out interval to the 11139 readout pulses send on to the emitter. We
furthermore extract a histogram of photon counts in the readout time
interval and compare it to a time interval of the same length where
no laser pulse was applied (see Fig.\ref{fig:SingleShot}b)). We extract
the readout fidelity following the definition in \cite{LucasHighFidSingleShotReadout}
and set a threshold of one detected photon for a state readout to
be detected as bright, while no detected photon corresponds to the
state to be dark. The readout error for a bright state to be detected
as dark is $\epsilon_{\text{B}}=0.45$, while the readout error for
a dark state to be detected as bright can be extracted from the dark
pulse measurement with additionally taking into account the \SI{1.1}{\%}
of residual population resulting from imperfect state initialisation,
which yields $\epsilon_{\text{D}}=0.08$.  \textcolor{black}{The latter is limited by the dark counts of the APD of about \SI{150}{cts/s}}. From the averaged state
readout error $\epsilon=\frac{1}{2}(\epsilon_{\text{B}}+\epsilon_{\text{D}})$
we calculate the total readout fidelity as $\mathcal{F}=1-\epsilon=0.74$.
This readout fidelity of \SI{74}{\%} surpasses the threshold of \SI{50}{\%}
above which a meaningful readout can be implemented, thus demonstrating
the potential of the $\text{SnV}^{-}$centre's strong cycling transitions.
The readout fidelity can be further improved by using superconducting
nanowire single photon detectors which have an improved dark count
rate of below \SI{1}{Hz} compared to 100-\SI{200}{Hz} of the APDs
used in our experiments. Furthermore, since our total collection efficiency
from bulk diamond is below 1\% there are straightforward measures
to improve the number of collected photons and thereby enabling high
fidelity single shot readout without the necessity of an aligned magnetic
field.

\section*{Discussion}

In this work, we thoroughly investigated the charge processes taking
place for $\text{SnV}^{-}$centres in diamond. We find a dominant
single photon electron capture process of single $\text{SnV}^{-}$centres
while being in the excited state as the reason for termination of
fluorescence under resonant excitation. \textcolor{black}{We propose a charge cycle
model, potentially universal to all G4V centres, which explains the charging and de-charging
in the presence of an additional divacancy defect.} Based on this model
we demonstrate a highly efficient \textcolor{black}{and rapid} charge initialisation exceeding
\SI{90}{\%} utilising a laser with a wavelength of \SI{445}{nm}
as a charge repump. The charge stabilised single $\text{SnV}^{-}$centres
are investigated in terms of the long term stability of the resonance
line with and without undergoing a full charge cycle. We find that
the charge stabilisation preserves very stable optical transitions
with deviations of the line centers well below the lifetime-limited
linewidth over the duration of one hour in both cases. Furthermore,
we probe the spin coherence of a charge stabilised $\text{SnV}^{-}$centre
and find a spin lifetime $T_{1}>$\SI{20}{ms} and a spin dephasing
time of $T_{2}^{*}$= \SI{5(1)}{\mu s} at a temperature of
\SI{1.7}{K}. As an outlook for further applications, we implement
a single shot readout scheme which yields a readout fidelity of \SI{74}{\%},
demonstrating the highly cycling spin conserving transitions even
in the case of a large angle of \SI{54.7}{\degree} between magnetic
field and the symmetry axis of the defect. It is straightforward to
increase the readout fidelity by implementing means to improve the
total collection efficiency of our optical setup such as utilising
optical antennas \cite{BecherOpticalAntenna}, micropillars \cite{EnglundTransformLimSnV},
solid immersion lenses, or photonic crystal structures \cite{LoncarSnVPhotCrystal,VuckovicSnVNanobeam}.
In summary, we have shown that the charge stabilisation protocol developed
in this work renders the $\text{SnV}^{-}$centre suitable for reliable
application in QIP, in which well defined charge states, \textcolor{black}{long-term}
stable optical resonances for the emission of highly indistinguishable
photons, long spin coherence times and efficient state readout are
absolutely crucial.

\section*{Methods}

\subsection*{Sample preparation}

\textcolor{black}{The predominantly investigated sample NI58 is an (001) electronic
grade bulk diamond, with the sample substrate being obtained by Element Six. It is specified to contain less than \SI{5}{ppb} (typically 0.1-\SI{1}{ppb}) of substitutional nitrogen, while the boron concentration is below \SI{1}{ppb}. The sample is homogeneously implanted with tin ions
at an implantation energy of \SI{700}{keV} and a fluence of 8x$10^{13}\frac{\text{Ions}}{\text{cm}^{2}}$.}
A subsequent annealing at \SI{2100}{\degree}C and \SI{7.7}{GPa}
is employed to reduce implantation damage. A detailed characterisation
of the sample can be found in \cite{NJPSpecInv}. \textcolor{black}{The second sample
BOJO\_001 is produced starting with a diamond substrate of the same type as used for NI58 and is investigated in the Supplementary Note 3. It is
produced by tin ion implantation at an implantation energy of \SI{700}{keV}
and 4 different fluences ranging from $10^{9}\frac{\text{Ions}}{\text{cm}^{2}}$
to $10^{12}\frac{\text{Ions}}{\text{cm}^{2}}$.} A subsequent annealing
\SI{2100}{\degree}C and \SI{8}{GPa} for \SI{2}{h} strongly reduces
implantation damage.

\subsection*{Experimental setup}

The measurements were conducted in a home built confocal microscope
with the sample being situated inside an closed cycle helium cryostat
(attodry2100, attocube systems AG) which operates at a base temperature
of \textasciitilde{}\SI{1.7}{K}. The sample is moved via a combination
of stepper (2x ANPx51, ANPz51, attocube systems AG) and scanner positioners
(ANSxy50, attocube systems AG). We use an NA 0.9 objective (MPLN100x,
Olympus) for photon collection.

The second wavelength for the fluorescence enhancement measurement
is retrieved from a supercontinuum laser source (SuperK FIANIUM FIU-15,
NKT Photonics) with \textasciitilde{}\SI{100}{ps} long pulses and
with a bandwidth of \SI{10}{nm}. The resonant laser fields are generated
from a side of fringe stabilised Dye cw laser (Matisse 2DS, Sirah
Lasertechnik GmbH) which can be locked to a wavemeter (WS6-200, HighFinesse
GmbH), while the laser light at \SI{445}{nm} is emitted by a diode
laser (Cobolt 06-MLD, H\"{u}bner Photonics). Pulses are carved from the
cw lasers using acousto optical amplitude modulators (AOM, AOMO 3200-146,
Crystal Technology ).

The timing of the pulse sequences for the electron capture, charge initialisation
and single shot readout experiments is controlled via an digital delay
generator (DG645, Stanford Research Systems).

The optical sidebands for the CPT experiments are introduced by an
electro-optical phase modulator (EOM, WPM-K0620, AdvR). The microwave
signals used to drive this modulator are emitted from an microwave
generator (mg3692c, Anritsu) and sent through a microwave amplifier
(ZHL-42+, Mini-Circuits). 

For the single shot readout, a second microwave generator (SG384,
Stanford Research Systems) is employed in order to implement two sidebands
with frequencies resonant with both SC transitions. The sidebands
are switched using two microwave switches (ZASW-2-50DRA+, Mini-Circuits).

\section*{Data availability}

The underlying data for this manuscript is openly available in Zenodo
at \newline http://doi.org/10.5281/zenodo.5561219. 

\section*{Code availability}
The evaluation algorithms
are available from the corresponding author upon reasonable request.

\subsection*{Acknowledgements}

\textcolor{black}{We thank Anna Maria Fuchs for significant contribution to reducing the computational overhead of the matrix density formalism and the enhancement measurement on the SiV ensemble.}
We thank Elke Neu and Richard Nelz for providing the additional microwave
source and for fruitful discussions. We thank Adam Gali and Jero Maze
for helpful theoretical insight and discussions on the charge cycle
of colour centres in diamond.

This research received funding from the European Union\textquoteright s
Horizon 2020 research and innovation programme under Grant Agreement
No. 820394 (ASTERIQS), the EMPIR programme co-financed by the Participating
States and from the European Union\textquoteright s Horizon 2020 research
and innovation programme (Project No. 17FUN06 SIQUST and 20FUN05 SEQUME), the German
Federal Ministry of Education and Research (Bundesministerium f\"{u}r
Bildung und Forschung, BMBF) within the projects Q.Link.X (Contract
No. 16KIS0864) and QR.X (Contract No. 16KISQ001K) and the Deutsche
Forschungsgemeinschaft (DFG, German Research Foundation) -{}- Project-ID
429529648 -{}- TRR 306 QuCoLiMa (\char`\"{}Quantum Cooperativity of
Light and Matter\textquoteright \textquoteright ).

T.I. and M.H. acknowledge the Toray Science Foundation and the MEXT
Quantum Leap Flagship Program (MEXT Q-LEAP) (Grant No. JPMXS0118067395).

T.T. acknowledges support from MEXT Q-LEAP (Grant No. JPMXS0118068379)
and JSPS KAKENHI (Grant No. JP20K21096).

\section*{Competing interests}

The authors declare no competing interests.

\section*{Author contributions}

J.G. and D.He. designed and J.G. conceived the experiment. J.G., D.He.,
P.F. and C.B. discussed and designed the charge cycle model. T.I.
and T.T. prepared the main sample, D.R., J.G., D.He., D.Ha. and P.-O.
C. prepared the additional HPHT sample. M.H. and M.M. supervised the
HPHT sample preparation for the two samples. C.B. supervised the whole
project. The paper was written with input from all authors.

\subsection*{Materials and correspondence}

C.B. and J.G. should be addressed for correspondence and material
requests. Email: christoph.becher@physik.uni-saarland.de; j.goerlitz@physik.uni-saarland.de

\subsection*{Supplementary Information}

Supplementary Information for this paper is available at XXX.

\end{document}